\documentclass{article}

\usepackage{arxiv}

\usepackage[utf8]{inputenc} 
\usepackage[T1]{fontenc}    
\usepackage{hyperref}       
\usepackage{url}            
\usepackage{booktabs}       
\usepackage{amsfonts}       
\usepackage{nicefrac}       
\usepackage{microtype}      
\usepackage{mathtools}
\usepackage{graphicx}

\usepackage{doi}
\usepackage{amsmath, amssymb}
\usepackage{indentfirst}
\usepackage{tikz}

\usepackage[square, numbers]{natbib}

\title{SYMBA: Symbolic Computation of Squared Amplitudes in High Energy Physics with Machine Learning}

\author{ {\hspace{1mm}Abdulhakim Alnuqaydan}\\
	Department of Physics and Astronomy\\
	University of Kentucky, USA\\
	Department of Physics, College of Science
	\\Qassim University, KSA\\
	\texttt{aal700@uky.edu} \\
	\And
	{\hspace{1mm}Sergei Gleyzer} \\
	Department of Physics and Astronomy\\
	The University of Alabama, USA\\
	\texttt{sgleyzer@ua.edu} \\
	\And
	{\hspace{1mm}Harrison Prosper} \\
	Department of Physics\\
	Florida State University, USA\\
	\texttt{hprosper@fsu.edu}\\
	}

\date{}

\renewcommand{\headeright}{}
\renewcommand{\undertitle}{}

\hypersetup{
pdftitle={SYMBA: Symbolic Computation of Squared Amplitudes in High Energy Physics with Machine Learning},
pdfsubject={q-bio.NC, q-bio.QM},
pdfauthor={Abdulhakim Alnuqaydan, },
pdfkeywords={physics, Second keyword, More},
}

\begin{document}
\maketitle

\begin{abstract}

The cross section is one of the most important physical quantities in high-energy physics and the most time consuming to compute. While machine learning has proven to be highly successful in numerical calculations in high-energy physics, analytical calculations using machine learning are still in their infancy. In this work, we use a sequence-to-sequence model, specifically, a transformer, to compute a key element of the cross section calculation, namely, the squared amplitude of an interaction.  We show that a transformer model is able to predict correctly 97.6\% and 99\% of squared amplitudes of QCD and QED processes, respectively, at a speed that is up to orders of magnitude faster than current symbolic computation frameworks. We discuss the performance of the current model, its limitations and possible future directions for this work. \footnote{Code repository: \url{https://github.com/ML4SCI/SYMBAHEP}}

\end{abstract}

\section{Introduction}
 Most machine learning applications in high-energy physics focus on numerical data (see, for example, Refs.\,\citep{Albertsson2018, Bourilkov2019, 2019, Feickert2021}, while only a few studies have investigated the application of machine learning to symbolic data \citep{Butter2021}.  Working with symbolic data is a challenging task both for humans and machines. Hand manipulation of large symbolic expressions is error prone, hence the routine use of domain-specific symbolic manipulation software tools\,\citep{SHTABOVENKO2020107478, UHLRICH2021107928}. The question we address in this paper is whether it is feasible to encapsulate accurately the highly non-trivial symbolic manipulations encoded in these tools in a machine learning model? Our motivation is to amortize the time required to compute symbolic expressions by accepting an upfront cost in creating them, so that one can reap the benefits later of a symbolic calculation that is much faster than the software tools used to generate the symbolic data.

 We use a \emph{sequence-to-sequence} (\texttt{seq2seq}) model, specifically, a \emph{transformer} \citep{Vaswani2017},  to compute symbolically the square of the particle interaction amplitude, a key element of a cross section calculation. To the best of our knowledge, this is a first application of such models to this task. 

Cross sections are important quantities in high-energy physics because they link the real world of experimental physics with the abstract world of theoretical models. The calculation of a cross section can be an exceedingly complicated procedure, requiring many mathematical operations, including Lorentz index contractions, color factor calculation,  matrix multiplications, Dirac algebra, traces, and integrals. Moreover, the complexity of these operations increases dramatically with the number of final state particles. It is a testament to the high mathematical sophistication and the depth of domain knowledge of the developers of these symbolic manipulation tools that such analytical calculations have been automated (see, for example, \texttt{FeynCalc} \citep{SHTABOVENKO2020107478}, \texttt{CompHEP} \citep{CompHEP:2004qpa} or \texttt{MARTY} \citep{UHLRICH2021107928}). 
The squaring of amplitudes can result in  very long expressions and correspondingly long  computation times, an important challenge to overcome in practical applications. In this work, we address this challenge using symbolic machine learning and demonstrate a  proof-of-concept for the symbolic calculation of the key element of cross section calculations. 

The paper is organized as follows. In Sec.~\ref{seq:related_work} we note related work. In Sec.~\ref{sec:background} we provide the context for the work and define the relevant quantities involved in high-energy physics cross section calculations. We also introduce the \texttt{seq2seq} models, followed by the pertinent details of the datasets, transformer model and its training parameters in Sec.~\ref{sec:model}. We present our results in Sec.~\ref{sec:results}, followed by a discussion in Sec.~\ref{sec:discussion} and our conclusions in Sec.~\ref{sec:conclusions}

\section{Related Work}
\label{seq:related_work}
Sequence-to-sequence  models are the basis of natural language processing (NLP) systems that are frequently used to translate from one language to another\citep{https://doi.org/10.48550/arxiv.1806.00187}, summarize text \citep{10.1145/3437802.3437832} and even map images to textual summaries of them \citep{DBLP:journals/corr/abs-2101-10804}. 
Transformer-based models \citep{Vaswani2017} are now considered the state-of-the-art in NLP applications. The most well-known NLP models are: BERT (Bidirectional Encoder Representations from Transformers) \citep{https://doi.org/10.48550/arxiv.1810.04805} and GPT (Generative Pre-trained Transformer) \citep{https://doi.org/10.48550/arxiv.1412.2306}. These models have been used in medicine \citep{https://doi.org/10.48550/arxiv.1907.09538} \citep{https://doi.org/10.48550/arxiv.2103.10504} and DNA analysis \citep{Ji2020.09.17.301879}. 

 Recently, \texttt{seq2seq} models have been used to perform symbolic calculations in calculus and to solve ordinary differential equations symbolically, achieving excellent results \,\citep{Lample2019}.
 This model was able to predict correctly the symbolic solutions of integrals and found solutions to some problems that could not be completed using the the standard symbolic mathematics systems within a time limit orders of magnitude longer than the model execution time. 
 Sequence-to-sequence transformer models have also been used to infer the  recurrence relation of underlying sequences of numbers \citep{https://doi.org/10.48550/arxiv.2201.04600}. The authors used a transformer model to successfully find symbolic recurrence relations given the first terms of the sequence. Transformer models have also been used for symbolic regression \citep{Valipour2021}. 

 In physics, symbolic machine learning, such as symbolic regression, have been applied to classical mechanics problems \citep{Cranmer2019, Lemos2022,Udrescu2019}.  In Ref. \citep{Butter2021}, symbolic regression was used to extract optimal observables as an analytic function of phase space applied to Standard Model Effective Field Theory (SMEFT) Wilson coefficients using LHC simulated data. In our work, we take a further step and show that symbolic machine learning  relying on transformers can be applied to complex, realistic, calculations in high-energy physics such as the calculation of squared amplitudes.

\section{Background}
\label{sec:background}

We begin with an overview of the context of the work reported in this paper. More details can be found in \cite{book}, however, for the benefit of readers who are not high-energy physicists, we provide a concise description.

The Standard Model, one of the great intellectual achievements of the 20th century, describes all known elementary particles and their interactions through three of the four known fundamental forces, namely, the weak, electromagnetic, and strong forces (see, for example, \cite{book}).
The Standard Model is a Quantum Field Theory (QFT) specified in a mathematical expression called a Lagrangian. In QFT, the elementary particles are described in terms of quantum fields in space-time, where each type of particle is associated with a different field. The interactions among these particles are governed by fields, which are sometimes referred to as force carriers, whose details are precisely determined by the symmetries imposed on the Lagrangian. 

\begin{figure} 
    \centering
    \includegraphics[width=6cm]{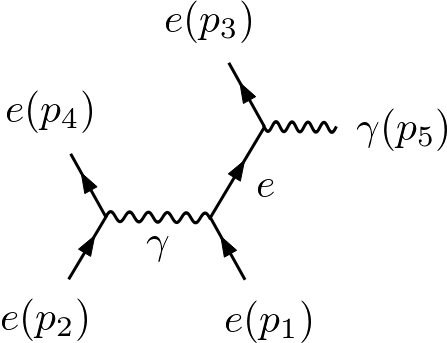}    \caption{A Feynman diagram of two incoming electrons $e(p_1)$ and $e(p_2)$ scattering into two electrons and a photon.}
    \label{fig:feyn_2to3}
\end{figure}

\begin{figure}
    \centering
    \includegraphics[width=16.9cm]{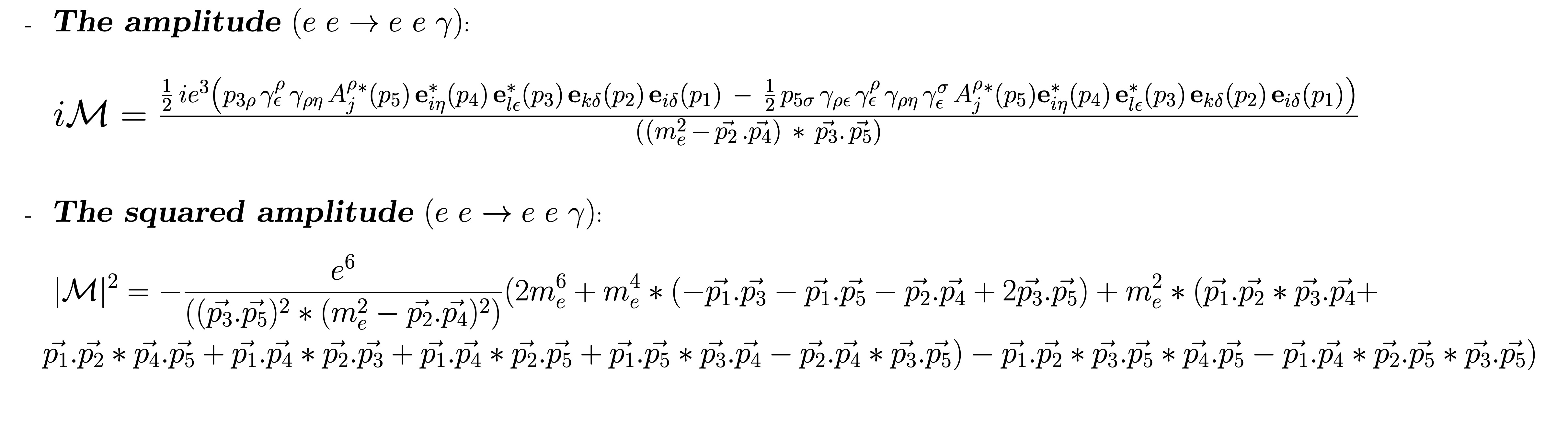}
    \caption{Amplitude and squared amplitude of the $e e \rightarrow e e \gamma$ scattering process.}
    \label{fig:ampl_sq}
\end{figure}

There is a well-defined procedure to extract from the Lagrangian all the possible particle interactions of interest, each associated with a mathematical quantity called an amplitude. These amplitudes can be represented by Feynman diagrams, as shown in Fig.~\ref{fig:feyn_2to3}.

Calculating the cross section, for example of the process depicted in Fig.~\ref{fig:feyn_2to3} and represented symbolically in  Fig.~\ref{fig:ampl_sq}, requires computing the squared amplitude and averaging and summing over the internal degrees of freedom of the particles. 

As noted above, these calculations are typically performed with domain-specific computer algebra frameworks, such as \texttt{FeynCalc} \citep{SHTABOVENKO2020107478}, \texttt{CompHEP} \citep{CompHEP:2004qpa} and \texttt{MARTY} \citep{UHLRICH2021107928}. 
Figure \ref{fig:ampl_sq} shows one of the ``shortest'' $2 \rightarrow 3$ quantum electrodynamic (QED) expressions after simplification, where in this process two incoming electrons $e(p_1)$ and $e(p_2)$ scatter into two electrons $e(p_3)$ and $e(p_4)$ and a photon $\gamma(p_5)$.  Typical expressions can have hundreds of terms and the computational time can become a major challenge, especially if higher-order amplitudes are included to render predictions more precise. 

The key insight in all the current uses of machine learning for symbolic applications is that many tasks can be viewed as a language translation problem. For example, a system that maps images to textual summaries of them can be viewed as translating from the language of images to a natural language. Likewise,  algebraic manipulation such as the mapping of amplitudes to their squared form can be conceptualized as a language translation task. Since this task maps one sequence of symbols to another it is natural to consider  \texttt{seq2seq} models.

\section{Model Description and Training} \label{sec:model}

\begin{figure}
    \centering
    \includegraphics[width=6cm]{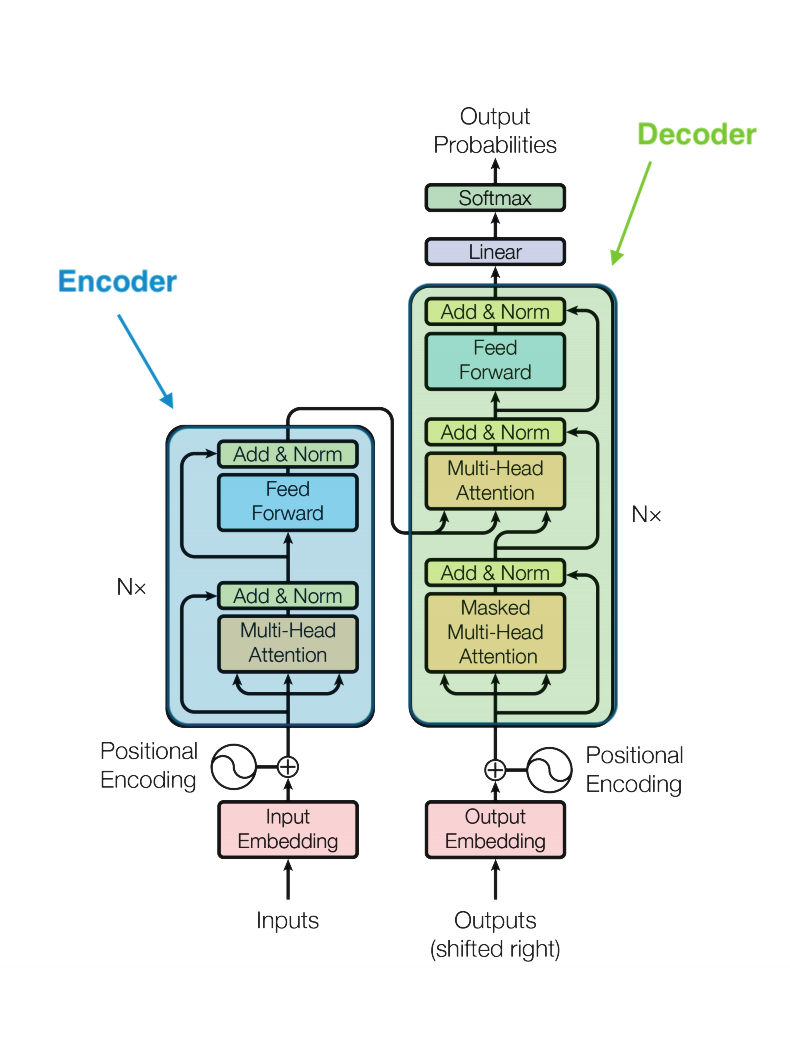}
    \caption{Transformer model architecture (reproduced from Ref. \citep{Vaswani2017}).}
    \label{fig:transformer}
\end{figure}

 In recent years, sequence-to-sequence transformer models have become one of the most powerful models for a variety of \texttt{seq2seq} applications \citep{Vaswani2017}.
Transformers use a mechanism called self-attention, whereby the model identifies and makes use of long-range dependencies among the symbols (or tokens) of the sequences. The architecture of the model consists of two parts: an \emph{encoder} and a \emph{decoder}, as shown in Fig. \ref{fig:transformer}.
The encoder first maps the input sequence to a vector in a  high-dimensional vector space in a process called \emph{embedding}, and encodes the position of each token in the sequence, an operation referred to as \emph{positional encoding}. Next, a measure of the degree to which a given token is related to other tokens, in a mechanism called multi-head attention, is calculated.
During the training of the model, the decoder takes the output vector from the encoder, which encodes information about the input sequence to the encoder, together with the encoded target sequence one token at a time and outputs a sequence also one token at a time.

\subsection{Datasets, Model and Training}

We use the symbolic computation program \texttt{MARTY} \citep{UHLRICH2021107928} to generate expressions for possible interactions in quantum electrodynamics (QED), the theory of the electromagnetic force, and quantum chromodynamics (QCD), the theory of the strong forces. For our proof-of-concept, we restrict the scope 2-to-2 and 2-to-3 particle tree-level processes.
All interactions involving off-shell and on-shell particles, anti-particles gauge bosons and triplet and anti-triplet color representations for QCD are included. Since it is possible for different amplitudes to yield the same squared expressions, we include such amplitudes in our dataset. Many of the QED and QCD interactions are non-trivial and lead to complicated expressions. All expressions are simplified with
the \texttt{Python} symbolic mathematics module \texttt{SymPy} \citep{10.7717/peerj-cs.103}, factorized by particle masses, and organized into a standard format (first the overall factors, then the terms of the numerator, and third the denominator). For practical reasons, we exclude expressions longer than $264$ tokens after simplification which excludes $5\%$ and $26\%$ of all QED and QCD expressions. This procedure yields 251,000 QED and 140,000 QCD expression pairs. The data are split into three sets: training, validation and test, 70\%, 15\% and 15\%, respectfully, and we choose a random sample of $500$ expression pairs (from the test set) to evaluate the performance of the trained model.

\begin{figure}
    \centering
   \includegraphics[width=7cm]{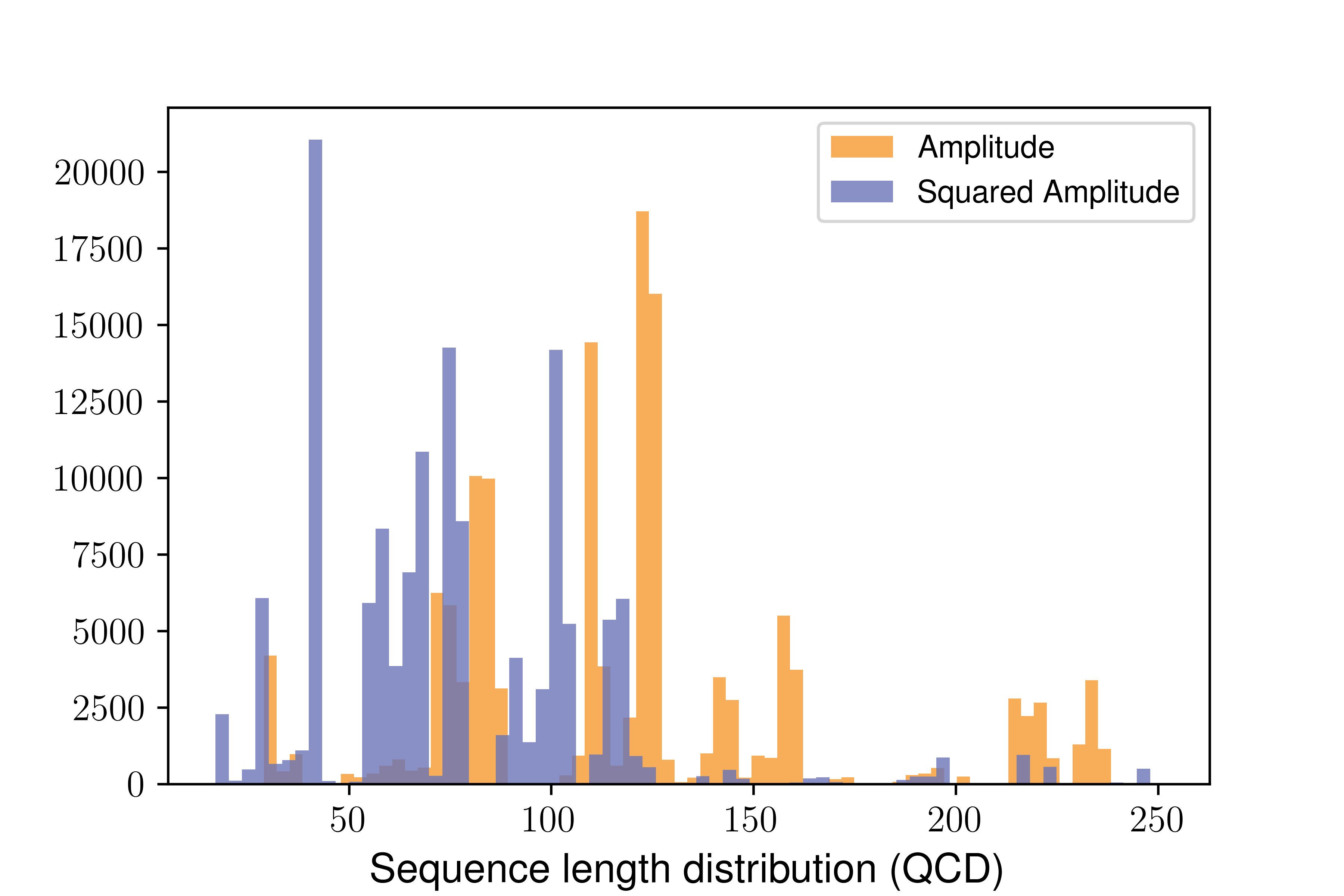}
    \includegraphics[width=7cm]{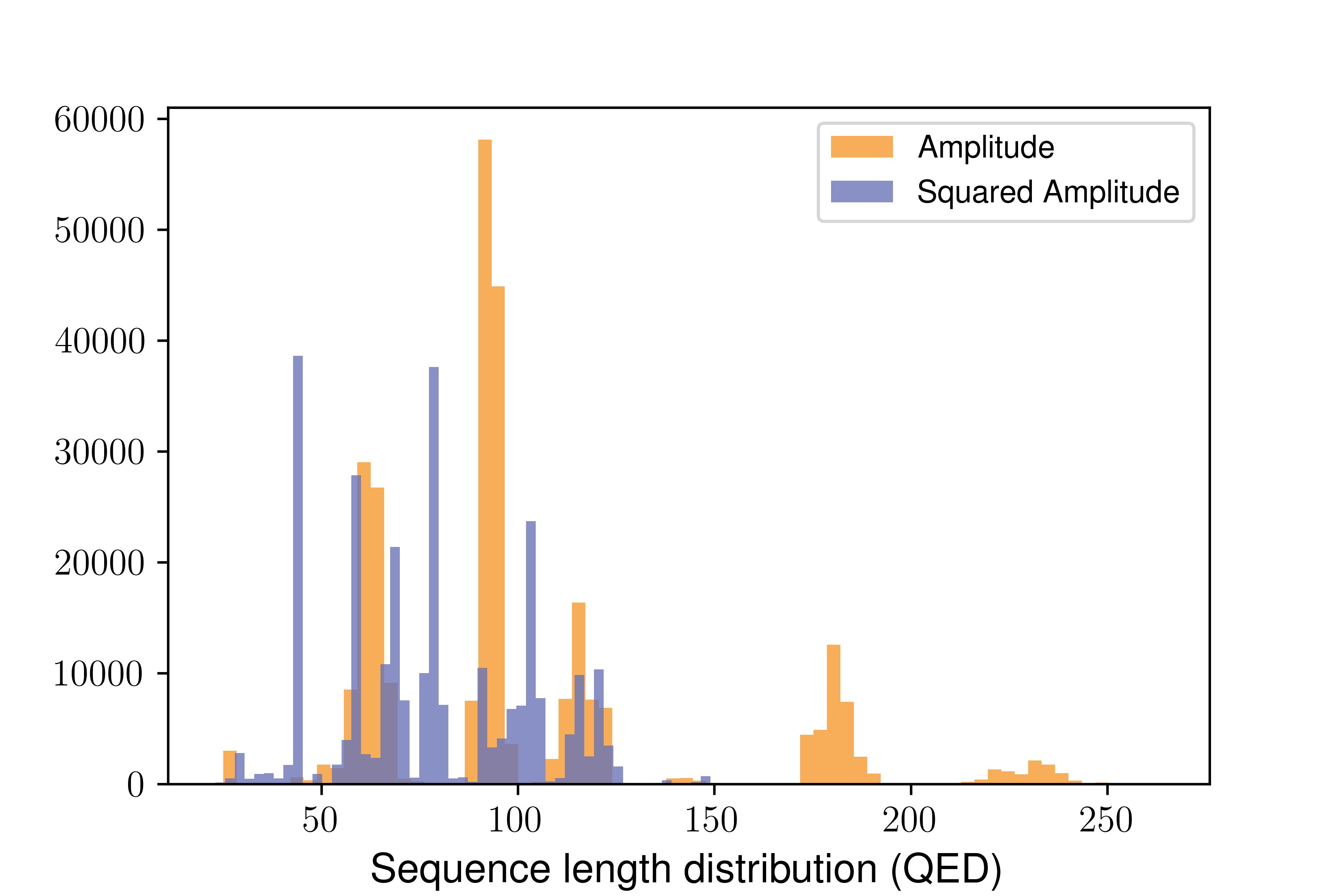}
    \caption{Sequence length distributions for QCD and QED sequences after expression simplification.}
    \label{fig:len}
\end{figure}

The \texttt{Tensorflow} \citep{tensorflow2015-whitepaper} package is used to map symbols to integers, specifically with the \texttt{TextVectorization}  class, which performs the tokenization, that is, the assignment of  an integer to each symbol and the padding of sequences to make them of equal length. Each sequence is then converted  to a vector built from these integers. The amplitudes are tokenized by operator (tensor) and its indices, while for squared amplitudes we tokenize them by each mass, product of momenta and numerical factor (for example, $4*m_e^2*\Vec{p_1}.\Vec{p_2}$ is three tokens) as there are a finite number of terms consistent with the physical dimension (in powers of mass) and conservation laws.

The transformer model is implemented using \texttt{Tensorflow 2.8} \citep{tensorflow2015-whitepaper} along with \texttt{Keras 2.8} \citep{chollet2015keras} without structural modifications. The model has $6$ layers and 8 attention-heads, with $512$ embedding dimensions and $8192$-$16384$ latent dimensions. We use \textit{sparse categorical crossentropy} as the loss function, the \texttt{Adam} optimizer \citep{https://doi.org/10.48550/arxiv.1412.6980} with a learning rate of $10^{-4}$ and a batch size of $64$. The training was performed for $50$-$100$ epochs on two CASCADE-NVIDIA V100 GPUs which took about $12$-$24$ hours. 

\section{Results} \label{sec:results}

The accuracy of the predicted symbolic expressions is assessed by taking a random sample of 500 amplitudes (from the test set) that have not been seen by the transformer model and predicting their squared amplitudes. We use three distinct metrics to assess the model accuracy.

\begin{figure}[t]
    \centering
    \includegraphics[width=8cm]{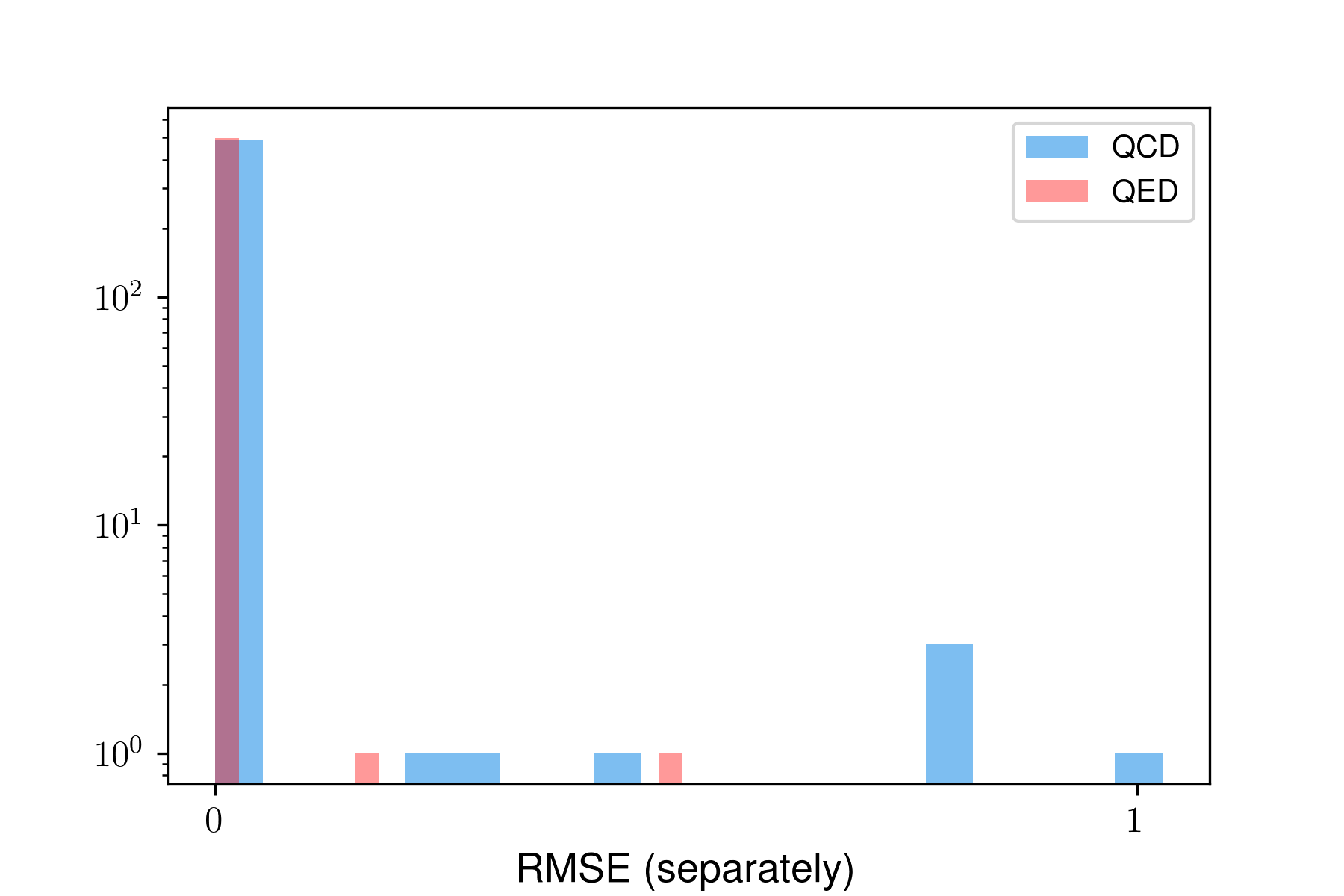}
    \includegraphics[width=8cm]{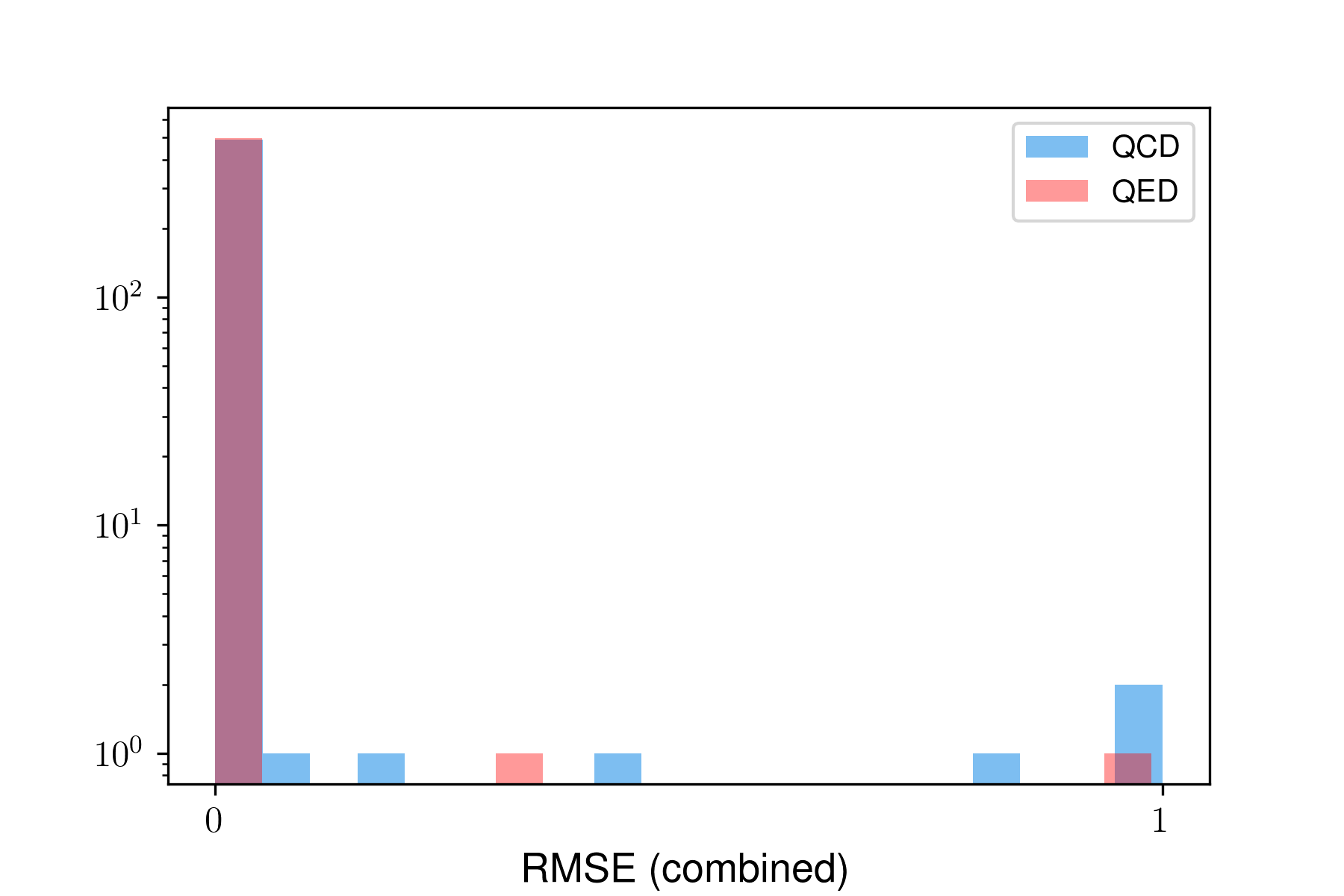}
    \caption{(left) RMSE distributions for QCD and QED expressions when models are trained separately for QCD and QED. (right) RMSE distributions when a single model is trained using the combined QCD and QED dataset.}
    \label{fig:nu_err}
\end{figure}

\begin{enumerate}
    \item \textbf{Sequence Accuracy:} the percentage of predicted symbolic expressions that \textit{identically} match the correct expression.\\
    \item \textbf{Token Score:} a measure of the number of tokens (symbols) predicted correctly in the correct location in the sequence. 
    \begin{equation*}
        Token \ Score = \frac{n_{c} - n_{ex}}{n_{act}}  \quad \%,
    \end{equation*}
    where $n_{c}$ is the number of tokens predicted correctly, $n_{ex}$ is the number of extra tokens that the model predicts (if any), and $n_{act}$ is the number of tokens in the correct sequence.\\
    
    \item \textbf{Numerical Error:} the relative difference between the values of the predicted and the correct expressions when using random numbers between \{0, 100\} for the momenta in the squared amplitude expression.
    \begin{equation*}
        Numerical \ Error = \frac{x_{act} - x_{pred}}{x_{act}},
    \end{equation*}
    where $x_{act}$ is the numerical value of the actual (correct) squared amplitude,  $x_{pred}$ is the predicted numerical squared amplitude. Repeating this process $100$ times and varying the random numbers each time, we compute the Root-Mean-Square-Error (RMSE) whose distributions are shown in Fig.\ref{fig:nu_err}
    
\end{enumerate}

\begin{table}[t]
	\caption{Amplitude-squared amplitude Model results}
	\centering
	\begin{tabular}{lllll}
		\toprule
	              
		\quad & Training Size  & Sequence Acc. & Token Score & RMSE\\
		\midrule
		\textbf{QED (sequence)} &  251K  & 98.6\% &   99.7\%   & 1.3$\times 10^{-3}$ \\
		\\
		\textbf{QCD (sequence)}  &  140K    & 97.4\% &  98.9\%   & 8.8$\times 10^{-3}$
		\\
		\\
		\textbf{(QED + QCD) on QED}  &  391K   & \textbf{99.0\%} &   99.4\%   & 2.5$\times 10^{-3}$
		\\ 
		\\
		\textbf{(QED + QCD) on QCD}  &  391K   & \textbf{97.6\%} &  98.8\%  & 6.8$\times 10^{-3}$  \\
	\bottomrule
		
		\\ \textbf{QED (diagram)}  &  258K  &  \textbf{99.0\%}  & 99.7\%  & 9.3$\times 10^{-4}$  \\
       \\
       \textbf{QCD (diagram)}  &  142K  &  73.4\%  & 82.0\%  & 0.3 \\
	\end{tabular}
	\label{tab:table}
\end{table}

Table \ref{tab:table} summarizes the performance  of our models. For QED, the model trained on QED expressions, correctly and identically predicts the squared amplitude for $493$ out of $500$ ($98.6\%$) amplitudes. The remaining $7$ squared amplitudes differ by one or more tokens. The overall token scores is $99.7\%$, i.e., the model correctly predicts at least $99.7\%$ of the tokens on average. For QCD, the model trained on QCD expressions has a sequence accuracy of $97.4\%$ and the token score is $98.9\%$. The average numerical error is $1.3$$\times 10^{-3}$ for QED and $8.8$$\times 10^{-3}$ for QCD (excluding two anomalous results with RMSE > 10 in QCD only). The model trained on the combined QED and QCD dataset achieves a sequence accuracy of $99.0\%$ for QED with a token score of $99.4\%$ and an RMSE of $2.5$$\times 10^{-3}$. The combined model gives a sequence accuracy of $97.6\%$ and token score of $98.8\%$ and RMSE of $6.8$$\times 10^{-3}$ for QCD.

\section{Discussion} \label{sec:discussion}

It has been demonstrated that even for a mathematical problem as complicated as squaring an amplitude, averaging and summing of over the internal degrees of freedom of particles, and manipulating the result into a meaningful form, it is possible to encapsulate that domain knowledge in a transformer model. Our model can ``do the algebra'' with a sequence accuracy between 97.6\% and 99.0\%. The time of inference, on average, varies from one second to two seconds. The prediction time for QED is slightly longer (or similar) to the time taken by \texttt{MARTY} for the same calculation. For QCD, we find that our model is up to 2 orders of magnitudes faster than \texttt{MARTY} for some amplitudes, (especially those with anti-triplet color representations).

We observe, as it shown in  Table \ref{tab:table_data_size} and Table \ref{tab:table_len}, that the accuracy of the mapping is primarily affected by two factors: the amount of training data and the sequence length. The accuracy of the QED results is slightly higher than for QCD, which reflects the fact that the QED training sample is larger than the QCD sample and the QED sequences are shorter on average than those for QCD. This is especially true of the amplitude expressions, which are typically more complicated for QCD than for QED.

It is noteworthy  that when trained on the combined QED and QCD datasets, we obtain a better performance overall and a higher performance on the QED and QCD test sets separately. This indicates that our results are still data limited. Therefore, we should anticipate performance improvements if we are able to create and use larger datasets. There are several ways to achieve that. For example, additional tree-level processes can be included, such as 2-to-4 or 2-to-5, or by going beyond tree-level processes. Additionally, we can add predictions from theories of beyond the Standard Model (BSM) physics as long as the mathematical structure of these theories is similar to that of the Standard Model.

\begin{table}[t]
	\caption{Model performance on different sizes of QCD and QED  dataset}
	\centering
	\begin{tabular}{lllllll}
		\toprule
	              
		\textbf{QCD} & \textbf{Train on:} & $\nicefrac{1}{10}$ data & $\nicefrac{1}{5}$ data & $\nicefrac{1}{3}$ data & $\nicefrac{1}{2}$ data & full data (140K)\\
      
		\quad & \textbf{Sequence Accuracy:}  &  52.0\%  & 82.0\% &   91.0\%   & 94.0\% &  97.4\% \\
		\\
		\midrule
        \textbf{QED} & \textbf{Train on:} & $\nicefrac{1}{10}$ data & $\nicefrac{1}{5}$ data & $\nicefrac{1}{3}$ data & $\nicefrac{1}{2}$ data & full data (251K)\\
        
		\quad & \textbf{Sequence Accuracy:}  &  76.4\%  & 89.6\% &   95.8\%   & 98.0\% &  98.6\% \\
		\\
  
	\end{tabular}
	\label{tab:table_data_size}
\end{table}

\begin{table}[t]
	\caption{Model performance on different sequence lengths of QCD and QED dataset}
	\centering
	\begin{tabular}{llllll}
		\toprule
	              
		\textbf{QCD}& \textbf{Maximum sequence length:}  & $170$ tokens & $195$ tokens & full length ($256$ tokens) 
        \\
        
        \quad & Sample size: & 
     \scriptsize{123K} & \scriptsize{126K} & \scriptsize{140K}
     \\
		\quad & \textbf{Sequence Accuracy:}  &  99.6\% &   98.6\%   & 97.4\%  
        \\
        \\
       
        \midrule
		\textbf{QED}& \textbf{Maximum sequence length:} & $170$ tokens & $185$ tokens & full length ($264$ tokens) 
        \\
        
        \quad & Sample size: & 
     \scriptsize{218K} & \scriptsize{237K} & \scriptsize{251K}
     \\
		\quad & \textbf{Sequence Accuracy:}  & 99.0\%  &   98.8\%  & 98.6\%  
        \\
        \\
	\end{tabular}
	\label{tab:table_len}
\end{table}
The complexity of the transformer model increases quadratically with sequence length, which makes the training highly resource-intensive and affects the accuracy. This challenge can be addressed with variants of the basic transformer model that exhibit better scaling with sequence length \citep{https://doi.org/10.48550/arxiv.2007.14062} \citep{https://doi.org/10.48550/arxiv.2004.05150}. The length of the sequence can be further tuned by adjusting the tokenization process.  Balancing the sequence length and number of tokens is an important hyperparameter optimization step in this class of models that we leave to future work.

Another issue that needs to be addressed is how to identify when a predicted squared amplitude is anomalous when there is no access to the correct answer. We have seen that a single token error involving a quantity with a large value, such as the top quark mass, impacts the numerical result severely. It would be highly desirable to have a built-in solution rather than rely on human intervention by, for example, automating dimensional analysis (as each term should have the same dimension), or checking if the expression respects the conservation laws. While these practical solutions may be sufficient in some cases, a machine-learning solution whereby a confidence level can be assigned to the overall expression as well as on a per-token basis may be needed for this task. We leave this to future work. \\
\\

\begin{figure}
    \centering
    \includegraphics[width=14cm]{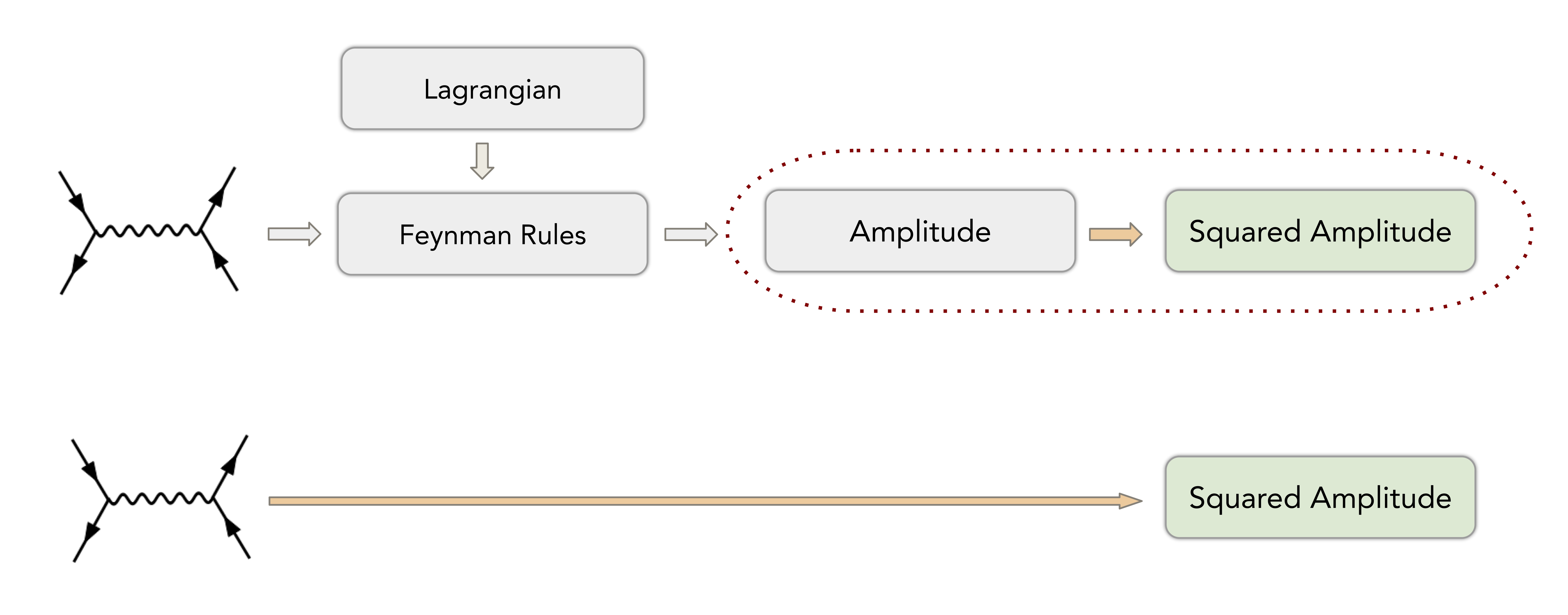}
    \caption{Amplitude to squared amplitude and Feynman diagram to squared amplitude workflows.}
    \label{fig:flow}
\end{figure}

As sequence-to-sequence models can be applied to all types of sequences, we can consider another interesting direction. Since Feynman diagrams themselves can be written down as sequences, we can map the Feynman diagram to the squared amplitude or to the cross section directly, as shown schematically in Fig. \ref{fig:flow}. A notable advantage of this is that the input Feynman diagram can be written by hand, if desired, without the need for a domain-specific tool to construct the amplitude. As a proof-of-concept of this idea, we test the transformer model on 2-to-3 QED and QCD processes following the same procedures as before. Remarkably, the QED Feynman diagram-based sequence model attains a sequence accuracy of $99.0\%$ and token accuracy of $99.7\%$ with average RMSE of 9.3$\times10^{-4}$ (see Table \ref{tab:table}). For QCD, the accuracy is much lower than for QED; the model for QCD attains a sequence accuracy of $73.4\%$ and a token accuracy of $82.0\%$ with average RMSE of 0.3 (excluding anomalous results with RMSE > 10). The current accuracy in QCD is lower compared to the models relying on amplitude sequence information, but this is expected as the number of input tokens is much smaller and may also indicate the need for a much larger
training dataset, which we shall consider in a future work. 

\section{Conclusion}
\label{sec:conclusions}
Our results demonstrate the ability of a symbolic deep learning model to learn the mapping between a particle interaction amplitude and its square to high accuracy despite the complexity of the mapping process. We observe that the accuracy strongly depends on the sequence length. The model has some limitations; however, the results obtained are sufficiently promising to motivate the search for further performance improvements that will be the focus of future work.

\section{Acknowledgements}

We thank the University of Kentucky Center for Computational Sciences and Information Technology Services Research Computing for their support and use of the Lipscomb Compute Cluster and associated research computing resources. Special thanks to Grégoire Uhlrich for technical help. 
This work was  supported in part by the U.S. Department of Energy (DOE) under Award No. DE-SC0012447 (SG) and supported
 in part by the U.S. Department of Energy Award No. DE-SC0010102 (HP).
This work was performed in part at Aspen Center for Physics, which is supported by National Science Foundation grant PHY-1607611.

\bibliographystyle{unsrtnat}
\bibliography{references}

\end{document}